\RequirePackage{lineno}
\documentclass[twocolumn,showpacs,preprintnumbers,prb]{revtex4}
\usepackage{graphicx,bm,amsmath,amssymb}

\def\gz{\ifmmode{Z\hskip -4.8pt Z}
    \else{\hbox{$Z\hskip -4.8pt Z$}}\fi}

\newcommand{\be}{\begin{equation}}
\newcommand{\ee}{\end{equation}}
\newcommand{\bea}{\begin{eqnarray}}
\newcommand{\eea}{\end{eqnarray}}

\begin{document}

\title{Scaling of conductance through quantum dots with magnetic field}
\author{I. J. Hamad}
\affiliation{Instituto de F\'{\i}sica Rosario. Facultad de Ciencias Exactas Ingenier\'{\i}a y Agrimensura, Universidad
Nacional de Rosario. Bv. 27 de Febrero
210 bis, 2000 Rosario, Argentina}
\author{C.~Gazza}
\affiliation{Instituto de F\'{\i}sica Rosario. Facultad de Ciencias Exactas Ingenier\'{\i}a y Agrimensura, Universidad
Nacional de Rosario. Bv. 27 de Febrero
210 bis, 2000 Rosario, Argentina}
\author{J.~A.~Andrade}
\affiliation{Centro At\'{o}mico Bariloche and Instituto Balseiro, Comisi\'{o}n Nacional
de Energ\'{\i}a At\'{o}mica, 8400 Bariloche, Argentina}
\email{aligia@cab.cnea.gov.ar}
\author{A.~A.~Aligia}
\affiliation{Centro At\'{o}mico Bariloche and Instituto Balseiro, Comisi\'{o}n Nacional
de Energ\'{\i}a At\'{o}mica, 8400 Bariloche, Argentina}
\email{aligia@cab.cnea.gov.ar}
\author{P. Roura-Bas}
\affiliation{Dpto de F\'{\i}sica, Centro At\'{o}mico Constituyentes, Comisi\'{o}n
Nacional de Energ\'{\i}a At\'{o}mica, Buenos Aires, Argentina}
\author{P.~S.~Cornaglia}
\affiliation{Centro At\'{o}mico Bariloche and Instituto Balseiro, Comisi\'{o}n Nacional
de Energ\'{\i}a At\'{o}mica, 8400 Bariloche, Argentina}
\email{aligia@cab.cnea.gov.ar}

\begin{abstract}
Using different techniques, and Fermi-liquid relationships, we calculate the
variation with applied magnetic field (up to second order) of the
zero-temperature equilibrium conductance through a quantum dot described by
the impurity Anderson model. We focus on the strong-coupling limit $U \gg
\Delta$ where $U$ is the Coulomb repulsion and $\Delta$ is half the
resonant-level width, and consider several values of the dot level energy $E_d$,
ranging from the Kondo regime $\epsilon_F-E_d \gg \Delta$ to the
intermediate-valence regime $\epsilon_F-E_d \sim \Delta$, where $\epsilon_F$
is the Fermi energy. We have mainly used density-matrix renormalization
group (DMRG) and numerical renormalization group (NRG) combined with
renormalized perturbation theory (RPT). Results for the dot occupancy and
magnetic susceptibility from DMRG and NRG+RPT are compared with the
corresponding Bethe ansatz results for $U \rightarrow \infty$, showing an
excellent agreement once $E_d$ is renormalized by a constant Haldane shift.
For $U < 3 \Delta$ a simple perturbative approach in $U$ agrees very well
with the other methods. The conductance decreases with applied magnetic 
field for dot occupancies $n_d \sim 1$ and increases for  $n_d \sim 0.5$
or $n_d \sim 1.5$ regardless of the value of $U$. We also relate the energy scale for 
the magnetic-field dependence of the conductance with the width of low energy peak in the spectral density of the dot.
\end{abstract}

\maketitle



\section{Introduction}

\label{intro}

In the last years an enormous amount of research in the field of nanoscience
has been devoted to the transport through semiconducting 
\cite{gold,cro,gold2,wiel,grobis,kreti,ama} and molecular 
\cite{liang,kuba,yu,leuen,parks,roch,scott,parks2,serge,vincent} quantum dots (QDs), and to
manifestations of the Kondo effect in these systems. The semiconducting QDs
are artificial atoms created in two-dimensional electron gases by a suitable
application of electrostatic voltages and are characterized by a high 
tunability of the parameters. By contrast in the molecular QDs, the molecule
itself can be changed.

Many of these systems are described by the Anderson model for a magnetic impurity 
with spin 1/2 
[Eq. \ref{ham} below], with constant hybridization $V$ between the impurity level 
and the conduction electrons, whose unperturbed density of states $\rho$,
can also be assumed constant. The main parameters of this model can be taken as
the energy $E_d$ of the level localized in the QD relative to the Fermi
energy, which we take as $\epsilon_F=0$, half the resonant level width 
$\Delta= \pi \rho V^2$ (in the literature sometimes the total width 
$\Gamma=2\Delta$ is used), and the Coulomb repulsion $U$. As we shall see, the
conduction electron band width $2D$ also plays a role, although a minor one. In the Kondo
regime, $-E_d \gg \Delta$ and $E_d +U \gg \Delta$, all quantities depend on a
single energy scale $T_K \sim D\sqrt{\rho J} \mathrm{exp}[-1/(\rho J)]$ with 
$J= 2V^2 U/[-E_d(E_d+U)]$,\cite{wilson} and the localized spin at the QD is
compensated by the conduction electrons resulting in a singlet ground state
and maximum conductance (unitary limit) at temperatures 
$T \ll T_K$.\cite{wiel,kreti}

Recent experiments have studied the scaling laws for the conductance through
one QD in the Kondo regime for small ($\ll T_K$) bias voltage $V_b$, temperature $T$ 
and applied magnetic field.\cite{grobis,kreti,scott} This stimulated further
theoretical work on the subject concentrated mainly on the effect of a
finite (although small) bias voltage, which is a tough non-equilibrium
problem. \cite{rinc,sela,rati,roura,bal,sca,mu,ng,note} While it would be
desirable to express the non-equilibrium properties in terms of
thermodynamic ones, this task seems possible only for a limited number of
coefficients in the expansion of the conductance.\cite{sca} 
Instead, at $T=0$ and at equilibrium ($V_b=0$), the
magnetic-field dependence of the conductance, characterized by a coefficient 
$c_B$ which was addressed theoretically recently,\cite{merker,cb,merker2}
can be expressed in terms of the magnetic susceptibility and the second
derivative of the dot occupancy with respect to the magnetic field,\cite{cb}
using an extension of the Friedel sum rule \cite{lang} to finite magnetic field for a
spin conserving impurity model,\cite{yoshi,he1} used before in magnetotransport
through quantum dots.\cite{lady,none,Cornaglia2005}

The effects of magnetic field in systems of one and several QDs have  
been studied before experimentally \cite{gold,kogan,zum,ama2,liu,sco2} and 
theoretically \cite{lady,none,costi,logan2,rosch,fuj2,ros2,hbo,wri,tosi,dias} 
using several methods.
The coefficient $c_B$ has been calculated using numerical renormalization
group (NRG) and a ``superperturbation'' approach \cite{mu,note} recently.\cite{merker,merker2}
Unfortunately, the results presented in Ref. \onlinecite{merker}, while exact at the symmetric point, are in general inaccurate because they miss a term proportional to the second derivative of the occupancy
with respect to the magnetic field \cite{cb} (the effect of this term is responsible for the difference between top and middle lines in Fig. \ref{cbu3} below ). Correct 
results were later presented,\cite{merker2} but only 
in the weak
coupling regime $U \le 3 \Delta$. For $U
\rightarrow \infty$, the coefficient $c_B$ was calculated using a slave-boson 
mean-field approximation (SBMFA).
\cite{cb} However, direct comparison with NRG results shows that the SBMFA
does not provide a reliable dot occupancy $n_d$,\cite{see} and since $n_d$ and its second
derivative with respect to the magnetic field enter the expression of $c_B$
[see Eq. (\ref{cbe})], the results of Ref. \onlinecite{cb} are only
qualitatively valid. 

On the other hand, for $U \le 3 \Delta$, the possibility to reach the Kondo
regime ($-E_d \gg \Delta$ and $E_d +U \gg \Delta$) is questionable and limited
in any case to the symmetric situation $E_d=-U/2$. In general, in the Kondo
regime, the spectral density presents two charge-transfer peaks at $E_d$ and 
$E_d+U$ of total width $4 \Delta$ and a Kondo peak at the Fermi energy of
total width $\sim 2 T_K$.\cite{pru,logan,capac} In the symmetric case $E_d=-U/2$ 
only one peak (at the Fermi level) is present for $U \le 3 \Delta$,\cite{note2} 
and as we shall discuss in this work, its half width at half maximum for $U = 3 \Delta$ is $0.54 \Delta$. 
The similarity of all energy scales and in particular the
large $T_K$ implies that one has to reach magnetic-field energies $g \mu_B B$
of the order of $U$ to obtain a splitting in the spectral density \emph{twice} 
$g \mu_B B$,\cite{none} which is a distinctive signature of Kondo 
physics.\cite{moore,costi,logan2,rosch,fuj2} For smaller magnetic fields the
splitting is smaller. A splitting roughly consistent with $2g \mu_B B$ was
observed in nonequilibrium transport through QDs.\cite{gold,kogan,zum,ama2}
More recent experiments obtain a somewhat larger splitting. This is probably
due to the fact that these experiments should not be interpreted in terms of
the equilibrium spectral density and non-equilibrium calculations are
necessary.\cite{none,ros2,hbo,wri} In any case, to detect clearly the splitting
one needs devices for which $T_K \ll U$.

Experimentally in semiconductor QDs a wide range of values of $U/\Delta$ are
possible. Typical values of $U$ are around 1-2 meV,\cite{gold,grobis,ama}
and $\Delta$ can vary between 10 to 200 $\mu$eV, leading to ratios $U/\Delta$
near 10 in the first experiments,\cite{gold} 2 and 4.5 in the two devices
studied by Kretinin \textit{et al.} \cite{kreti}, and larger than 50 in a
recent system of two QDs.\cite{ama} 
In contrast in molecular QDs the
expected order of magnitude of $U$ is $\sim 1$ eV for small molecules \cite{roch,serge}
or $\sim 0.1$ eV for large molecules.\cite{kuba}, while typical values of 
$\Delta$ are of the order of $\sim 1$ meV.\cite{roch,serge}
Thus, the ratio $U/\Delta$ is several orders of magnitude and in practice one can take 
$U \rightarrow \infty$. These facts indicate the need to extend the 
reliable calculations of $c_B$ (so far limited to the weak coupling regime $ U \leq 3 \Delta$)
to more realistic values.

In addition, the molecular QDs are characterized by a high
asymmetry of the coupling of the QD to the leads, and as a consequence, the
shape of the differential conductance $G=dI/dV_b$ near zero applied bias
voltage $V_b$ reproduces the spectral density of the Kondo peak.\cite{capac,note1}
This renders possible to relate the energy scale of $c_B$ with the width of the
measured zero-bias anomaly as we shall discuss.

In this work we calculate $c_B$ as a function of $E_d$ for several values of $U$ and in particular
in the strong-coupling limit $U \rightarrow \infty$. We also calculate the half width at 
half maximum $\Delta_\rho$ of the low-energy peak in the spectral density of the dot for 
for several values of $E_d$ and $U$. 
As $E_d$ increases from the symmetric case
$E_d=-U/2$, $c_B$ decreases, changes sign in the intermediate-valence regime $E_d \sim 0$ 
(where the occupancy $n_d \sim 0.6$) and becomes large and negative in the empty-orbital regime
$E_d > 0$. As a function of $n_d$, $c_B$ looks qualitatively similar for all values of $U$.

The paper is organized as follows. In Section II we present the model and equations that 
relate $c_B$ with thermodynamic quantities. In Section III we describe briefly the different methods used.
Section IV contains the results and Section V is a summary.

\section{Model and formalism}
\label{model}

The model is an Anderson impurity one in which a single localized level in a
QD is hybridized with two conducting leads,

\begin{eqnarray}
H=& \sum_{\nu k\sigma }\epsilon _{\nu k}c_{\nu k\sigma }^{\dagger }c_{\nu
k\sigma }+\sum_{\sigma }E_{d}n_{d\sigma }+Un_{d\uparrow }n_{d\downarrow } \nonumber \\
& +\sum_{\nu k\sigma }(V_{\nu k}d_{\sigma }^{\dagger }c_{k\sigma }
+\mathrm{H.c.})-g\mu _{B}Bs_{z},
\label{ham}
\end{eqnarray}
where $c_{\nu k\sigma }^{\dagger }$creates a conduction electron at the
lead $\nu $ with momentum $k$ and spin $\sigma $ in the conduction band, $%
d_{\sigma }^{\dagger }$ creates an electron in\ a localized level of the QD, $%
n_{d\sigma }=d_{\sigma }^{\dagger }d_{\sigma }$, and $s_{z}=(n_{d\uparrow
}-n_{d\downarrow })/2$. By considering appropriate linear combinations of
the electrons of both leads, the model is mapped into a single-channel
model with resonant-level half-with at half-maximum $\Delta =\pi \sum_{\nu
k}|V_{\nu k}|^{2}\delta (\omega -\epsilon _{k})$, which we assume
independent of energy $\omega $.

At zero temperature ($T=0$) and at equilibrium (bias voltage $V_{b}=0$), the
contribution to the conductance of each spin for a given magnetic field 
$G_{\sigma }(B)$ is proportional to the corresponding density of states of
the dot level $\rho _{d\sigma }(\omega ,B)$ at the Fermi energy $\omega =0$.\cite{capac,meir} 
In turn, this quantity is related to the occupancy for the
corresponding spin by the Friedel sum rule \cite{lang} 
generalized to finite $B$ \cite{yoshi,he1,lady}

\begin{equation}
\rho _{d\sigma }(0,B)=\frac{\sin ^{2}(\pi n_{d\sigma })}{\pi \Delta },
\label{rhos}
\end{equation}
where for simplicity we denote as $n_{d\sigma }$ the expectation value of
the corresponding operator. This allows to express the change of conductance
by an applied magnetic field in terms of the occupancies

\begin{equation}
\frac{G(B)}{G(0)}=\frac{\rho _{d\uparrow }(0,B)+\rho _{d\downarrow }(0,B)}{%
\rho _{d\uparrow }(0,0)+\rho _{d\downarrow }(0,0)}.  \label{g1}
\end{equation}%
Expanding $n_{d\sigma }$ up to second order in $B$ one has \cite{cb}

\begin{equation}
n_{d\sigma }(B)=\frac{n_{d}}{2}+\frac{\chi B}{g\mu _{B}}\sigma +\frac{%
\partial ^{2}n_{d}}{\partial B^{2}}\frac{B^{2}}{4}+O(B^{3}),  \label{nd}
\end{equation}%
where $n_{d}=n_{d\uparrow }+n_{d\downarrow }$, $\chi $ is the magnetic
susceptibility, $\sigma =1$ (-1) for spin up (down) and the quantities in
the second member except $B$ are evaluated at $B=0$. 

From these equations, and defining $c_{B}$ and $T_{0}$ 
(an energy scale of the order of $T_K$)
by \cite{merker,cb,merker2}

\begin{eqnarray}
\frac{G(B)}{G(0)} &=&1-c_{B}\left( \frac{g\mu _{B}B}{T_{0}}\right) ^{2}, 
\notag \\
\chi  &=&\frac{\left( g\mu _{B}\right) ^{2}}{4T_{0}},  \label{gb}
\end{eqnarray}%
one obtains

\begin{eqnarray}
c_{B} &=&\frac{\pi ^{2}}{16}(1-c^{2})
-c\frac{\pi }{2}\left( \frac{T_{0}}{g\mu _{B}}\right) ^{2}\frac{\partial ^{2}n_{d}}{\partial B^{2}},  \notag \\
\text{with }c &=&\cot \left( \frac{\pi n_{d}}{2}\right).   \label{cbe}
\end{eqnarray}

\section{Methods}
\label{meth}

As explained in the previous section, the calculation of the coefficient 
$c_B$ that describes the magnetic-field dependence of the conductance
at equilibrium and zero temperature, reduces to the calculation of the dot occupancy
$n_d$, the magnetic susceptibility $\chi$ and the second derivative
of the occupancy with magnetic field $\partial ^{2}n_{d}/\partial B^{2}$.
We have used four methods to calculate these quantities.

\subsection{Density-matrix renormalization group}
\label{dmrg}

We used density-matrix renormalization group (DMRG) \cite{dmrg} to solve the 
impurity Anderson model for $U/\Delta=3$, 8 and also for the infinite $U$ case. 
The band was discretized using a Wilson chain\cite{wilson}, with a discretization parameter $\Lambda$. 
This presents benefits that have already been introduced in the literature.\cite{ Nishimoto, Weichselbaum, Silva}
In the range of parameters considered, the results do not change significantly when increasing 
the number of sites $N$ beyond a certain value $N^*(\Lambda)$, as we have numerically verified for several values of $\Lambda$.  We found that setting $\Lambda=4$, $N=50$, and retaining $m=600$ states was enough to assure convergence, being the truncation error (the weight of the neglected states in the density matrix) $10^{-6}$ in the worst case, reassuring the reliability of the calculation. The DMRG results in this paper correspond to these values of $\Lambda$, $N$ and $m$.

As a consequence of this discretization, the density of
conduction states at the Fermi level $\rho_\Lambda(0)$ decreases with respect to 
the continuum limit $\Lambda \rightarrow 1$. We have calculated 
$\rho_\Lambda(0)$ numerically for the Wilson chain without dot, and determined 
the hybridization $V_\Lambda$ of the dot with the first site of
the chain from the condition $\pi \rho_\Lambda(0) V_\Lambda^2=\Delta$.

We have chosen a half band width $D=100 \Delta$ since the ratio $D / \Delta$ is very
large in real systems.

We have calculated the ground state energy and the occupancy of the impurity
for different values of the magnetic field $B$, which was applied to the
impurity site only, as in Eq. (\ref{ham}).
For the calculation of the susceptibility and $\partial ^{2}n_{d}/\partial B^{2}$, 
the general criterion used was to calculate the ground-state energy and magnetization at the dot 
for ten different values of 
the magnetic field $B$ such that $B_{min}<B<B_{max}$, where $g \mu_B B_{min}=0.005 T_{K}^0$, 
and $B_{max} \approx 150 B_{min}$. 
The curves were smooth and very well approximated by linear or quadratic polynomials, depending on the case. 
The susceptibility was calculated both as the derivative
of the magnetization of the dot with respect to $B$ fitting the curve with a straight line through the origin, 
and as the second derivative of the ground state energy with respect to $B$ fitting with a quartic polynomial 
and extracting the coefficient of the quadratic term. In all cases the agreement was complete.

\subsection{Numerical renormalization group plus renormalized perturbation theory}
\label{nrg}

We have used the standard Numerical renormalization group (NRG)  
\cite{krishna,bulla} to calculate the occupancy $n_d$ and its second derivative with respect 
to the magnetic field $\partial ^{2}n_{d}/\partial B^{2}$. The latter was obtained calculating
$n_d$ for nearly 10 different magnetic fields in the interval $0 < g\mu_B B < \widetilde{\Delta}$,
where $\widetilde{\Delta}$ is defined below and fitting the points with a $B^2$ dependence.

We used a discretization parameter 
$\Lambda=3.5$ and truncated the spectrum, after the fifth iteration, keeping
up to 2000 states. In contrast to the DMRG calculations, we take $D= 10 \Delta$ and do the
numerical calculations using a hybridization $V_\Lambda=\sqrt{A_\Lambda} V$ between
the dot and the first conduction site in the Wilson chain, where $\Delta= \pi V^2/(2D)$ and 

\be
A_\Lambda=\frac{\Lambda+1}{2 \Lambda-2}{\rm ln}\Lambda.
\ee
This expression has been suggested to obtain the correct Kondo temperature in the Kondo limit.\cite{krishna,campo}.

For the  magnetic susceptibility $\chi$ one has the problem that in the Kondo regime 
(large $U$ and $\Delta \ll |E_d|, E_d+U$) 
it oscillates as the iterations increase (lowering the temperature).
One way to solve this problem is to use $\Lambda \gg 1$ and average 
over several realizations of the logarithmic grid,\cite{mac} but we obtained
only a moderate improvement. A better method is to use the full density matrix 
within the NRG, although still in the strong coupling limit, values of the Wilson ratio $R$
above 2 and about 3\% larger than the exact ones were obtained.\cite{mac}
Here we have obtained renormalized parameters $\widetilde{E}_d$, $\widetilde{\Delta}$ and $\widetilde{U}$ 
that describe the low-energy physics,
following the procedure explained by Hewson {\it et al.} in Ref. \onlinecite{hom}.
The only difference is that we interpret that in Eq. (42) of that paper, the 
first member refers to $\widetilde{\Delta}_\Lambda$ the renormalized $\Delta$ for $\Lambda \neq 1$,
which is related to $\widetilde{\Delta}$ by $\widetilde{\Delta}_\Lambda=A_\Lambda \widetilde{\Delta}$.
From the renormalized parameters, the susceptibility is obtained accurately using 
renormalized perturbation theory (RPT) to second order in $\widetilde{U}/(\pi \widetilde{\Delta})$.\cite{he1}
The result is

\be
\chi  =(g\mu _{B})^{2}\widetilde{\rho }_{d}(0)R/2,  \label{xi}
\ee
where the Wilson ratio and the quasiparticle spectral density at the dot are  

\bea
R=1+\widetilde{U}\widetilde{\rho }_{d}(0), \label{r} \\
\widetilde{\rho }_{d}(\omega )=\frac{\widetilde{\Delta }/\pi }{(\omega 
-\widetilde{E}_{d})^{2}+\widetilde{\Delta }^{2}}.  \label{rhor}
\eea

Using this procedure we obtain for example for $U=10 \Delta$ and $E_d=-U/2$, a Wilson ratio $R=1.9939$, 
while the exact result from Bethe ansatz (see Section \ref{bethe}) is 1.9957 and NRG with the full density 
matrix gives 2.024.\cite{mac}

The quasiparticle spectral density is modified by the renormalized interaction $\widetilde{U}$.\cite{he1}
To calculate the width of the resulting renormalized spectral density we use ordinary
perturbation theory (PT) \cite{yos,hor} to second order in $\widetilde{U}/(\pi \widetilde{\Delta})$,
taking $\widetilde{E}_d$ as the effective dot energy. Since even for $U \rightarrow \infty$, 
$\widetilde{U}/(\pi \widetilde{\Delta})$ is of the order of 1 (see Table \ref{tabuinf}), the second order results are
accurate enough. To illustrate the consistency of the procedure, if one uses
directly PT for $U= 3\Delta$ in the symmetric case $E_d=-U/2$, one obtains a half width at 
half maximum $\Delta_\rho$ of the low-energy peak in the spectral density of the dot, only 
about 2\% higher ($\Delta_\rho=0.558 \Delta$) than
the result $\Delta_\rho=0.545 \Delta$ that comes determining first the renormalized parameters
by NRG and then using PT.
As $T_0$, $\Delta_\rho$ is also of the order of the Kondo temperature $T_K$ in units
for which the Boltzmann constant $k_B=1$.

\subsection{Bethe ansatz}
\label{bethe}

As a check to the results obtained with the methods described above, we have calculated
the occupancy and magnetic susceptibility for $U \rightarrow \infty$ and several
values of a shifted dot energy $E_d^*$ at $T=0$ using integral analytical
expressions obtained with the Bethe ansatz in Ref. \onlinecite{wt}
and as a particular case of a more general model in Ref. \onlinecite{abp}.
Some tricks to deal with singularities in the integrals were used, which are explained in the appendix of Ref. 
\onlinecite{anda}. Unfortunately, to obtain an analytical exact expression for the second derivative
of the occupancy with magnetic field $\partial ^{2}n_{d}/\partial B^{2}$, is very
difficult and is beyond the scope of the present paper. 
This precludes to give exact results for $c_B$ out of the symmetric case $E_d=-U/2$.

In the symmetric case, V. Zlati\'{c} and B. Horvati\'{c} have shown that 
the exact results for the  magnetic susceptibility and quasiparticle weight
$z=\widetilde{\Delta}/\Delta$ can be expressed as a convergent power series.\cite{zla}
We have evaluated this series using a FORTRAN program in quadruple precision.
From these results we have derived the renormalized parameters and the Wilson ratio
for comparison.

\subsection{Interpolative perturbative approximation}
\label{ipa}

For small values of $U/ \Delta$ one may use the interpolative perturbative approximation (IPA) which
is a modification of the PT approach to second order in $U /\Delta$  
modified to 
reproduce exactly the atomic limit $U/\Delta \rightarrow +\infty $.\cite{kk,ipa1,ipa2,levy}
In addition, the on-site term is split as 
$\sum_{\sigma }E_{d}n_{d\sigma }= 
\sum_{\sigma }E_d^{\rm eff}n_{d\sigma }+\sum_{\sigma }(E_{d}-E_d^{\rm eff})n_{d\sigma }$,
where the second term is included in the perturbation, and $E_d^{\rm eff}$ 
is determined to optimize the results.
In particular, in Ref. \onlinecite{kk}, $E_d^{\rm eff}$ was determined to satisfy
the Friedel sum rule. 
Results for the conductance through a QD using the IPA \cite{pro} agree with more recent ones 
using the finite temperature density matrix renormalization group method.\cite{maru}

Here we use the spin dependent version in which 
$E_{d \sigma}^{\rm eff}$ depends on spin and is determined self-consistently for each spin
to satisfy the corresponding Friedel sum rule Eq. (\ref{rhos}).\cite{none} 
Comparison with exact results without magnetic field shows that
the IPA provides very accurate values for the occupancy $n_d$ as a function of $E_d$
for $U /\Delta \leq 6$. The maximum deviation is less than 1 \% for $U /\Delta=6$.\cite{anda}

\section{Results}
\label{res}

In this section, we provide results of the coefficient $c_B$ of the magnetic-field
dependence of the conductance [see Eq. (\ref{gb})] and the energy scale $T_0$ 
as a function on the energy level of the QD $E_d$ which is easily controlled by a gate voltage 
in transport experiments. Since $T_0$ is determined by the magnetic susceptibility
$\chi$ [see Eq. (\ref{gb})] which is not accessible in transport experiments, we also provide
a relation between $T_0$ and the half width at half maximum of the spectral density
$\Delta_\rho$. In experiments with high asymmetry in the coupling between the dot and both leads
and in the Kondo regime, the differential conductance as a function of bias voltage $G(V_b)=dI/dV_b$ 
directly represents the Kondo peak in the spectral density.\cite{capac,note1} 
For less asymmetry and entering in the intermediate-valence regime, the half width at half maximum 
of the zero-bias peak in $G$ times the electric charge $e$ ($\Delta_G$) increases 
to about 1.6 $\Delta_\rho$ but remains of the same order of magnitude for $U \rightarrow \infty$.\cite{capac}

We focus our study in three values of $U/\Delta$: 3, which is the largest value for which
previous NRG results were reported,\cite{merker2}, $U \rightarrow \infty$ which corresponds
to the molecular QDs, and an intermediate case $U/\Delta=8$ which is a reasonable value for 
semiconducting QDs in the Kondo regime.

\subsection{$U/\Delta=3$}
\label{r3}

\begin{figure}[t]
\includegraphics[width=8.cm]{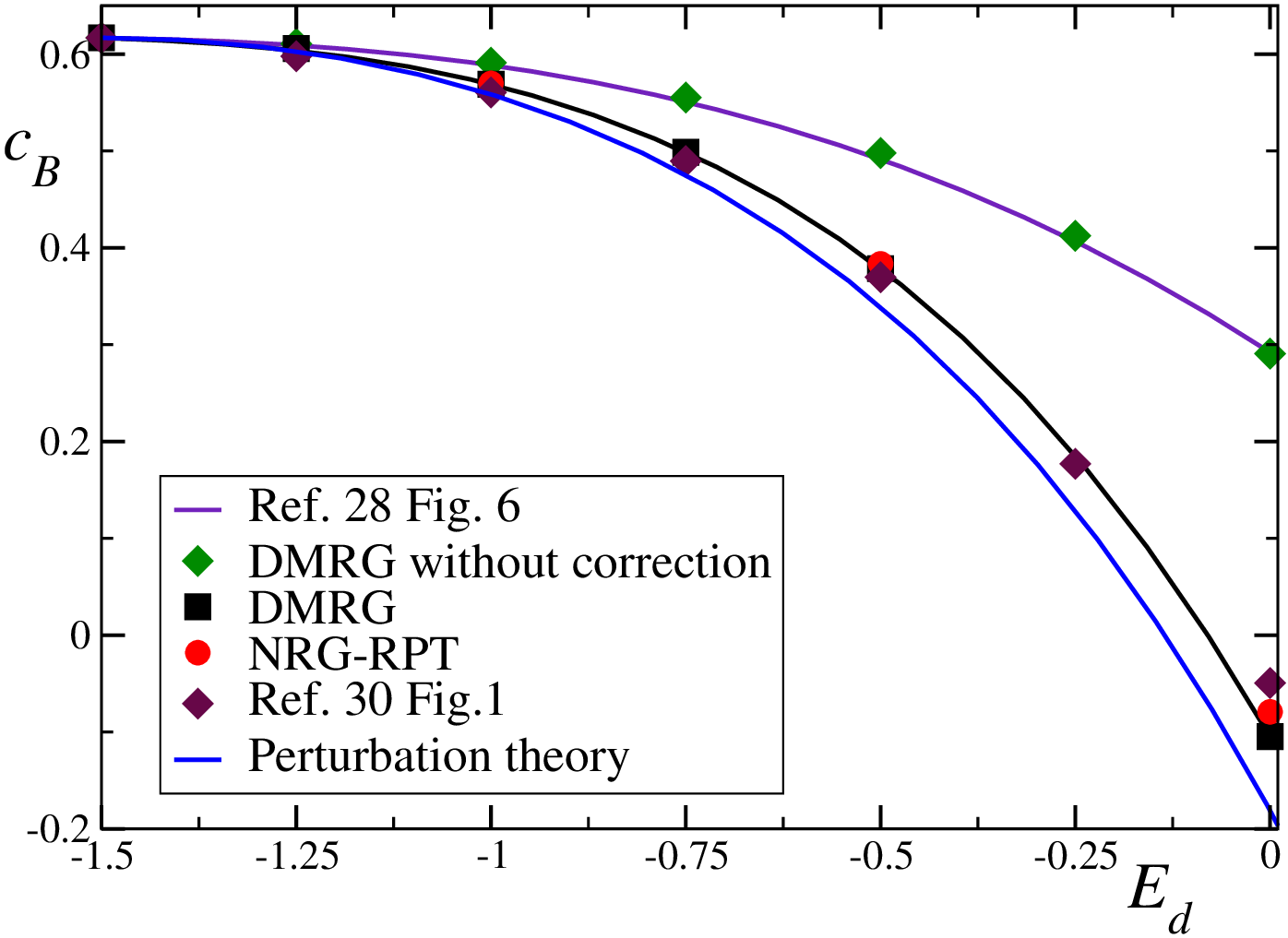}
\caption{(Color online) $c_B$ vs $E_d$ for $U=3 \Delta$ obtained by different methods.}
\label{cbu3}
\end{figure}

In Fig. \ref{cbu3} we represent the coefficient of the magnetic-field dependence of the conductance 
[see Eq. (\ref{gb})] as a function of the on-site energy for $U=3 \Delta$ and three different methods.
Because of a particular electron-hole symmetry (see Section II.C.1 of Ref. \onlinecite{vau}), 
$c_B (E_d)= c_B (-U-E_d)$, or in other
words, for $ -U \leq E_d \leq -U/2$, $c_B$ is the specular image around $E_d=-U/2$ of the results 
represented in Fig. \ref{cbu3}.
Our DMRG (NRG + RPT) results were obtained in intervals of $\Delta/4$ ($\Delta/2$) between the symmetric
case $E_d=-U/2$ and $E_d=0$. These results agree between them and also with the corrected NRG ones
of Merker {\it et al.}.\cite{merker2} To compare with these results, we have digitalized the results 
of $c_B^\prime$ given in Fig. 1 of Ref. \onlinecite{merker2} and used the relation

\be
\frac{c_B^\prime}{c_B}=\left(\frac{T_0^{\rm sym}}{T_0}\right)^2=\left(\frac{\chi}{\chi^{\rm sym}}\right)^2,  
\label{cbp}
\ee
where the superscript ``sym'' refers to the symmetric case $E_d=-U/2$. There is a small discrepancy 
for $E_d=0$, where the results of Merker {\it et al.} lie slightly above ours. This might be due to 
the fact that their results were calculated directly from the conductance, while ours use only
thermodynamic quantities and are expected to be more accurate. In fact we have also 
calculated $c_B$ from the conductance derived from the spectral density calculated 
with the full-density-matrix NRG for a few points, but the results showed
some deviations from the results obtained with DMRG and NRG+RPT using Eq. (\ref{cbe}).

We also show in Fig. \ref{cbu3} the result of $c_B$ within DMRG including only the first term 
of $c_B$ [neglecting the second negative term proportional to
$\partial ^{2}n_{d}/\partial B^2$ in Eq. (\ref{cbe})] 
and compared with the corresponding 
result in Fig. 6 of Ref. \onlinecite{merker}. 
There is again a good agreement with the results of Ref. \onlinecite{merker2} indicating a 
coincidence (except for deviations imperceptible in the figure) between the corresponding results 
for the occupancy $n_d$ in the range of $E_d$ studied.  Although not shown in the figure, the first term of $c_B$ 
(which depends only on the occupancy $n_d$) within NRG+RPT and IPA also coincides 
with the results shown, because all these methods provide accurately values for $n_d$.
Instead the full result requires the calculation of the magnetic susceptibility $\chi$
and $\partial ^{2}n_{d}/\partial B^2$ for which the IPA fails at large $U$ as explained below.

The correction due to the second derivative of the 
occupancy becomes significant in the intermediate-valence regime, in particular at $E_d=0$, where
it leads to a change of sign in $c_B$ [for all for all parameters considered in this paper, 
the first (second) term in Eq. (6) is positive (negative)]. The effect of this correction (or in general
the difference between different approximations for $c_B$) is not so evident in
the curve $c_B^\prime$ vs $E_d$ because the magnetic susceptibility decreases strongly 
as $E_d$ enters in the intermediate-valence regime (particularly for large $U$) and then 
the values of $|c_B^\prime|$ are largely reduced with respect to $c_B$ [by a factor 5.5
according to Eq. (\ref{cbp}) for $E_d=0$]. The different approaches lead to the exact result 
$c_B=c_B^\prime=\pi^2/16 \approx 0.617$ in the symmetric case $E_d=-U/2$,\cite{merker,cb,merker2,ca} 
and for large enough $U$ the values of $c_B^\prime$ lie near zero for $E_d=0$, even neglecting
$\partial ^{2}n_{d}/\partial B^2$, because of the factor $(T_0^{\rm sym}/T_0)^2$.

The values of $c_B$ obtained using IPA lie slightly below those of the other methods (in 
$c_B^\prime$ vs $E_d$ the maximum deviation in near 0.02). We have verified that the occupancy 
is very well reproduced by the IPA (with an underestimation of the order of 1 \%) in agreement
with a previous comparison with Bethe-ansatz results.\cite{anda} Therefore, the first term (positive 
in the range of $E_d$ shown) of $c_B$ in Eq. (\ref{cbe}) is well reproduced. However, the remaining  
negative term is overestimated
due to an underestimation of the magnetic susceptibility $\chi$ 
(overestimation of $T_0$) by 6\% and an overestimation of 
$\partial ^{2}n_{d}/\partial B^2$ that reaches of the order of 10\% for $E_d$ near $-\Delta$.
The accuracy of the IPA depends strongly on the perturbation parameter $U/(\pi \Delta)$. 
For example in the symmetric case $E_d=-U/2$, the IPA result for $\chi$ is below the Bethe ansatz results
by 15\% for $U=4 \Delta$ and by only 1.4\% for $U=2 \Delta$.

One disadvantage of $c_B$ with respect to $c_B^\prime$ is that the energy scale used
as a reference, $T_0$, depends on $E_d$ [see Eq. (\ref{gb})] and one needs to know two numbers
for each $E_d$ ($c_B$ and $T_0$) to describe the magnetic-field dependence of the conductance.
An advantage is that $T_0$ is of the order of the Kondo temperature, which in turn
is proportional to the width $2 \Delta_G/e$ of the zero-bias peak in the conductance 
(which is experimentally accessible) and to the 
width $2 \Delta_\rho$ of the spectral density of the dot state. All these quantities
are of the order of the quasiparticle level width $\widetilde{\Delta}$. In Table \ref{tabu3} we
give values obtained from RPT+NRG of these quantities, except $\Delta_G$. This width is addressed in 
Ref. \onlinecite{capac} where it is shown that in the Kondo regime for strongly asymmetric
leads (as usual in molecular QDs) $\Delta_G=\Delta_\rho$.\cite{note1}

\begin{table}[h]
\caption{\label{tabu3} Effective parameters, half width of the spectral density $\Delta_\rho$ 
and ratio $\Delta_\rho/T_0$ obtained from NRG+RPT for $U=3 \Delta$ and several values of
$E_d$. }
\begin{ruledtabular}
\begin{tabular}{llllll}
$E_d/\Delta$ & $\widetilde{\Delta}/\Delta$ & $\widetilde{E}_d/\widetilde{\Delta}$ & 
$\widetilde{U}/(\pi \widetilde{\Delta})$ & $\Delta_\rho/\widetilde{\Delta}$ & $\Delta_\rho/T_0$ \\ \hline
-1.5  & 0.639 & 0     & 0.738 &  0.835 & 0.944 \\
-1    & 0.671 & 0.196 & 0.732 &  0.848 & 0.885 \\
-0.5  & 0.754 & 0.421 & 0.716 &  0.883 & 0.766 \\
0     & 0.845 & 0.700 & 0.698 &  0.930 & 0.581 \\
\end{tabular}
\end{ruledtabular}
\end{table}

The Bethe-ansatz result of $\widetilde{\Delta}$ for $E_d= -1.5 \Delta$ is 0.6522, slightly larger than 
the RPT+NRG result 0.639 tabulated. This is due to the fact that for this calculation
we used a finite half band width $D=10 \Delta$ that affects the Kondo temperature. 
However the Wilson ratio 1.738 is almost the same as the exact one 1.741.
The ratio $\Delta_\rho/T_0$ evolves from near 1 in the symmetric case $E_d=-U/2$ to near
1/2 for $E_d=0$.
From the data given in Table \ref{tabu3} and Fermi-liquid relations one can obtain the 
occupancy

\be
n_d=1-\frac{2}{\pi} \arctan \left(\frac{\widetilde{E}_d}{\widetilde{\Delta}}\right),
\ee
which coincides in general within 1 \% with the corresponding result obtained directly with
NRG.

\subsection{$U \rightarrow \infty$}
\label{rinf}

\begin{figure}[h]
\vspace{1cm}
\includegraphics[width=8.cm]{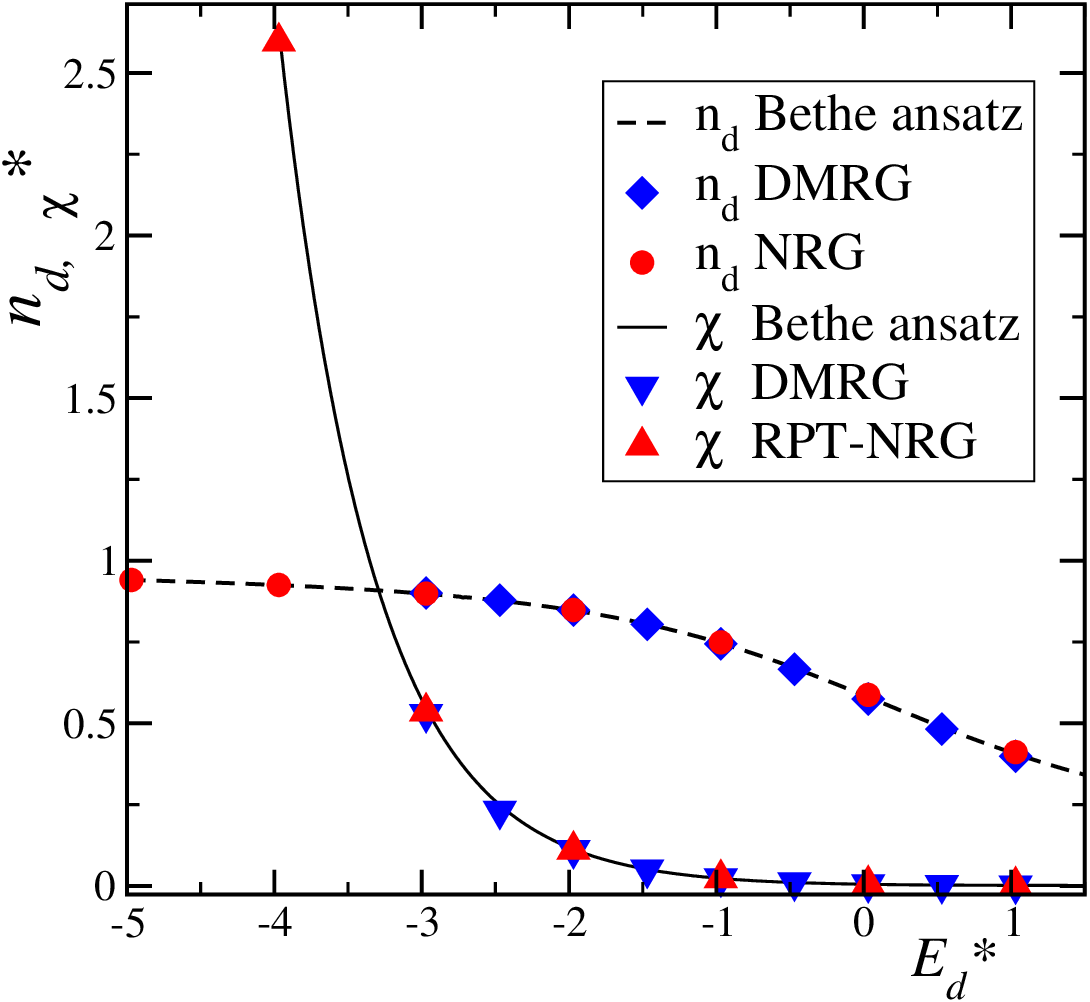}
\caption{(Color online) occupancy and scaled magnetic susceptibility $\chi^*=\chi \Delta/[100(g \mu_B)^2]$ 
for $U \rightarrow \infty$ as a function of the shifted dot energy for different techniques.}
\label{chin}
\end{figure}

Experimentally, the on-site energy of the dot level $E_d$ is controlled by the gate voltage
and determined by capacitance effects of all applied gate voltages \cite{ama,capac,haug,park}
(see for example the supplementary material of Ref. \onlinecite{ama}).
The position of the Coulomb blockade edges and the related charge-transfer peaks in the spectral density 
of the dot level are determined by a shifted energy $E_d^*$ which contains the effects
of a renormalization due to the hybridization with the leads.\cite{capac,hald}
This shift is significant when both, the half band width $D$ and the Coulomb repulsion $U$ are much
larger than $\Delta$, as in many realistic systems. In particular for $U \rightarrow \infty$, 
the calculation by Haldane based on poor man's scaling gives
\be
\delta = E_d^*-E_d = \frac{\Delta}{\pi} \text{ln} \left( \frac{D}{\alpha \Delta} \right),
\label{shift}
\ee
where $\alpha \sim 1$. Experimentally, $E_d^*$ is accessible but not $E_d$. For example 
at temperatures above the Kondo temperature, a maximum in the equilibrium conductance
takes place for $E_d^*=0$, or for finite $E_d^*$ Coulomb edges appear for bias voltages
such that $V_b= \pm E_d^*/e$.\cite{capac}

In order to compare the DMRG results for the occupancy $n_d$ and magnetic susceptibility $\chi$
for fixed $E_d$ with the corresponding analytical Bethe-ansatz (in which
$D \rightarrow \infty$ is assumed) as a function of $E_d^*$, we have to determine
the shift $\delta_{\rm DMRG}$. We obtain that {\em both} functions $n_d$ and $\chi$
shifted by {\em the same} $\delta_{\rm DMRG}=1.53 \Delta$ coincide with the corresponding
Bethe-ansatz results, as shown in Fig. \ref{chin}. 
This result is consistent with Eq. (\ref{shift}) which gives $\delta = 1.47 \Delta$
for $D=100 \Delta$ and $\alpha=1$.
The same happens with the 
NRG+RPT results using $\delta_{\rm NRG}=1.00 \Delta$.
For large negative values of $E_d/\Delta$ the occupancy flattens near 1, and the susceptibility 
increases strongly due to the exponential dependence of the Kondo temperature $T_K$ on 
$E_d^*$ and the fact that $\chi$ is proportional to $1/T_K$ in the Kondo regime. 

\begin{figure}[t]
\includegraphics[width=8.cm]{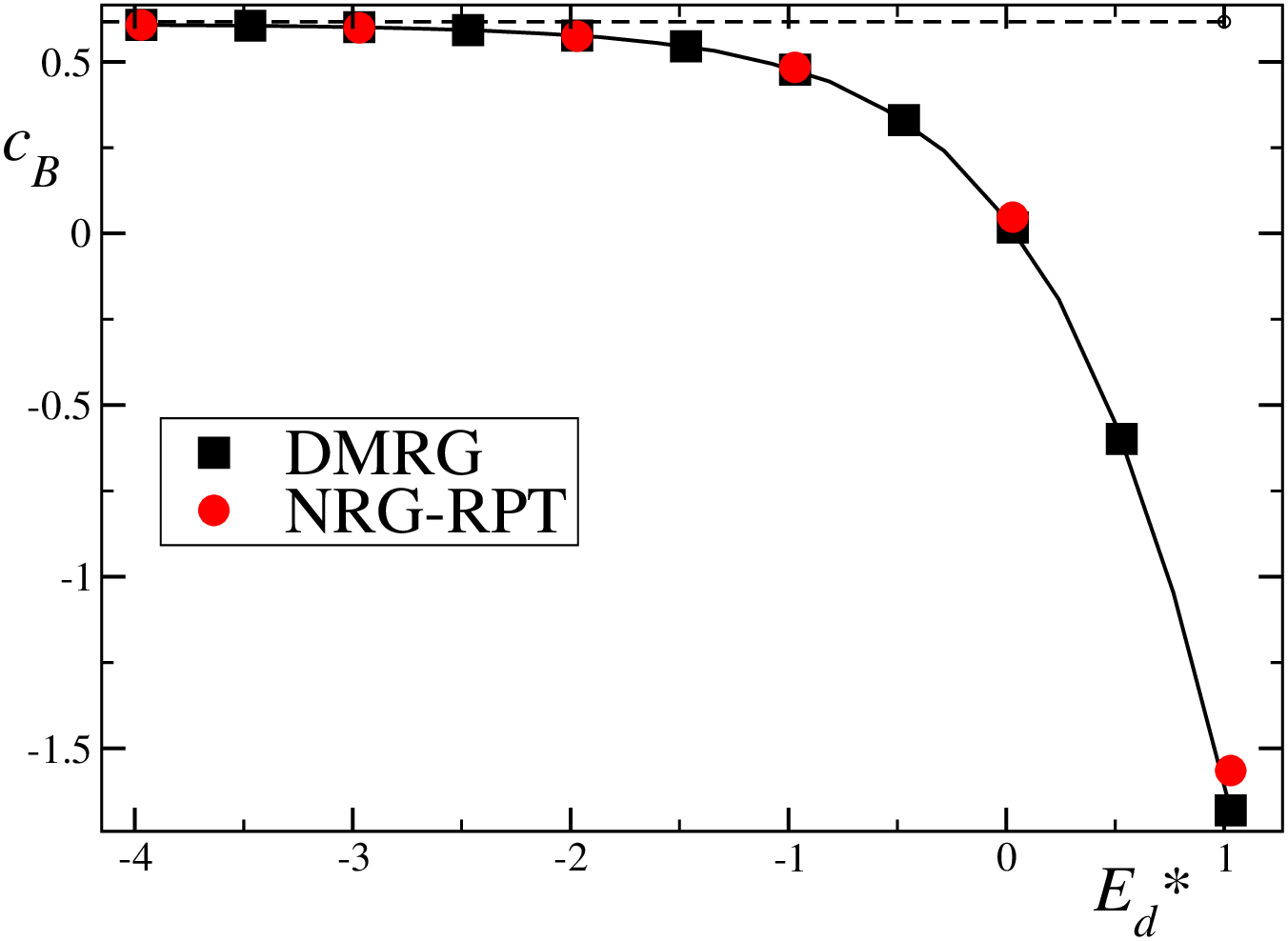}
\caption{(Color online) $c_B$ vs $E_d^*$ for $U \rightarrow \infty$ obtained by DMRG and NRG+RPT.}
\label{cbuinf}
\end{figure}

Once the shifts $\delta$ are determined, we can represent $c_B$ as a function of $E_d^*$ 
using both methods. The results are shown in Fig. \ref{cbuinf}. For $E_d^*/\Delta < -2$, as the 
occupancy $n_d > 0.84$ and the system is in the Kondo regime, $c_B > 0.5$ indicating a 
small to moderate deviation from the value $(\pi/4)^2 \approx 0.617$ of the symmetric case.
In this region, the correction due to the second (negative) term of $c_B$ in Eq. (\ref{cbe}) is below 1 \% 
Instead, for $E_d^*/\Delta > -2$, $c_B$ decreases rapidly and changes sign for $E_d^*/\Delta \approx 0$,
in the middle of the intermediate-valence region.

\begin{table}[h]
\caption{\label{tabuinf} Effective parameters, half width of the spectral density $\Delta_\rho$ 
and ratio $\Delta_\rho/T_0$ obtained from NRG+RPT for $U \rightarrow \infty$ and several values of
$E_d$. }
\begin{ruledtabular}
\begin{tabular}{llllll}
$E_d/\Delta$ & $\widetilde{\Delta}/\Delta$ & $\widetilde{E}_d/\widetilde{\Delta}$ & 
$\widetilde{U}/(\pi \widetilde{\Delta})$ & $\Delta_\rho/\widetilde{\Delta}$ & $\Delta_\rho/T_0$ \\ \hline
-6 & $2.51 \times 10^{-4}$ & 0.0937 & 1.009 &  0.706 & 0.892 \\
-5 & $1.21 \times 10^{-3}$ & 0.118 & 1.014 &  0.705 & 0.886 \\
-4 & $5.79 \times 10^{-3}$ & 0.160 & 1.025 &  0.703 & 0.873 \\
-3 & 0.0270 & 0.243 & 1.054 &  0.699 & 0.838 \\
-2 & 0.115 & 0.416 & 1.136 &  0.693 & 0.740 \\
-1 & 0.356 & 0.766 & 1.317 &  0.716 & 0.530 \\
0 & 0.640 & 1.338 & 1.594 &  0.793 & 0.286 \\
\end{tabular}
\end{ruledtabular}
\end{table}

In Table \ref{tabuinf} we list the renormalized parameters and ratios of different quantities proportional 
to $T_K$ obtained with NRG+RPT. As it is expected in the Kondo regime $-E_d \gg \Delta$ and $E_d +U \gg \Delta$,
the Kondo temperature depends exponentially on $E_d$ for $E_d/\Delta < -2$. However, in this regime, particularly
for $E_d/\Delta = -3$, the ratios $\Delta_\rho/\widetilde{\Delta}$ 
and $\Delta_\rho/T_0$ are almost
constant, near 0.7 and 0.85 respectively. Instead, entering the intermediate valence regime, 
$\Delta_\rho/\widetilde{\Delta}$ slightly increases and $\Delta_\rho/T_0$ strongly decreases.

\subsection{$U/\Delta=8$}
\label{r8}

\begin{figure}[h]
\vspace{0.5cm}
\includegraphics[width=8.cm]{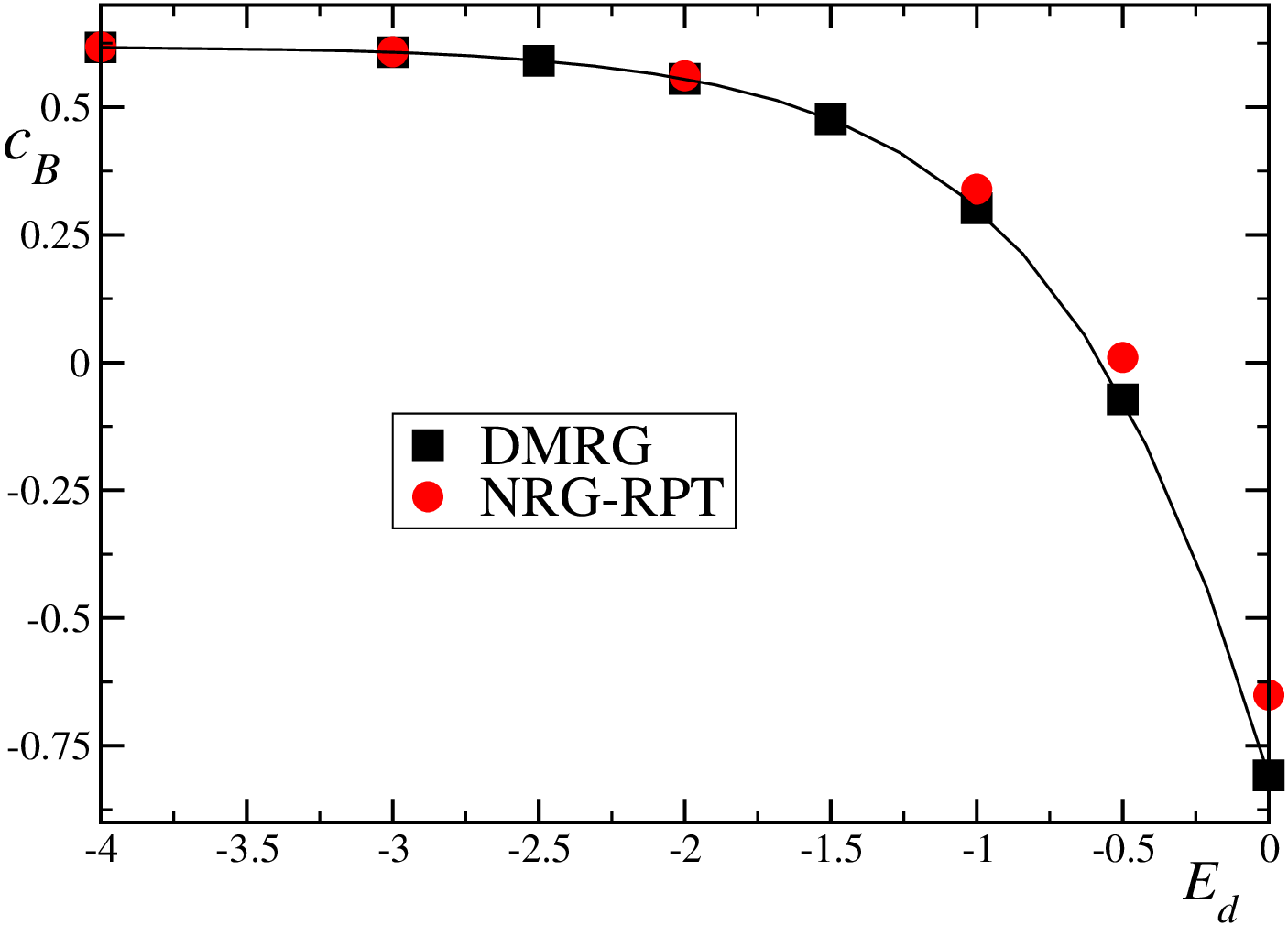}
\caption{(Color online) $c_B$ vs $E_d$ $U=8 \Delta$ obtained by DMRG and NRG+RPT.}
\label{cbu8}
\end{figure}

In Fig. \ref{cbu8} we show the $c_B$ for $U=8 \Delta$ obtained by DMRG and NRG+RPT.
While the results of both methods coincide in the Kondo regime ($E_d  \leq -2 \Delta$ 
in the figure) there are some deviations in the intermediate valence regime. This
is due to the fact that for finite $U$ the Haldane shift $\delta$ depends on 
$E_d$, being 0 in the symmetric case $E_d=-U/2$, positive for $E_d > -U/2$ and larger
for larger $D$. Therefore, the DMRG results calculated with $D$ 10 times larger, correspond 
to larger $E_d^*$ in general, but for both techniques $E_d^*=E_d$ in the symmetric case.

\begin{table}[h]
\caption{\label{tabu8} Effective parameters, half width of the spectral density $\Delta_\rho$ 
and ratio $\Delta_\rho/T_0$ obtained from NRG+RPT for $U=8 \Delta$ and several values of
$E_d$. }
\begin{ruledtabular}
\begin{tabular}{llllll}
$E_d/\Delta$ & $\widetilde{\Delta}/\Delta$ & $\widetilde{E}_d/\widetilde{\Delta}$ & 
$\widetilde{U}/(\pi \widetilde{\Delta})$ & $\Delta_\rho/\widetilde{\Delta}$ & $\Delta_\rho/T_0$ \\ \hline
-4 & 0.120 & 0     & 0.985 &  0.715 & 0.902 \\
-3 & 0.143 & 0.101 & 0.987 &  0.718 & 0.895 \\
-2 & 0.235 & 0.247 & 1.004 &  0.724 & 0.845 \\
-1 & 0.457 & 0.510 & 1.040 &  0.759 & 0.699 \\
-0.5 & 0.609 & 0.515 & 1.060 &  0.802 & 0.573 \\
0 & 0.746 & 0.977 & 1.081 &  0.857 & 0.430 \\
\end{tabular}
\end{ruledtabular}
\end{table}

Table \ref{tabu8} displays the renormalized parameters and ratios $\Delta_\rho/\widetilde{\Delta}$, 
$\Delta_\rho/T_0$ obtained with NRG+RPT. As in the case $U \rightarrow \infty$, these ratios
are nearly constant in the Kondo regime (although the variation is more pronounced as before)
$-E_d \gg \Delta$ and $E_d +U \gg \Delta$, while in the intermediate valence regime, 
$\Delta_\rho/\widetilde{\Delta}$ increases and $\Delta_\rho/T_0$ decreases markedly (although
not so strongly as for $U \rightarrow \infty$).
The Bethe ansatz result of $\widetilde{\Delta}/\Delta=0.1326$ for $E_d/=-4 \Delta$ 
is about 10 \% larger than the NRG+RPT result. As for $U= 3 \Delta$ this difference is due to the 
effect of the finite band width $2D$ of the conduction band on the Kondo temperature, and 
practically does not alter the Wilson ratio.  

\subsection{$c_B$ as a function of occupancy}
\label{cboc}

\begin{figure}[h]
\vspace{0.5cm}
\includegraphics[width=8.cm]{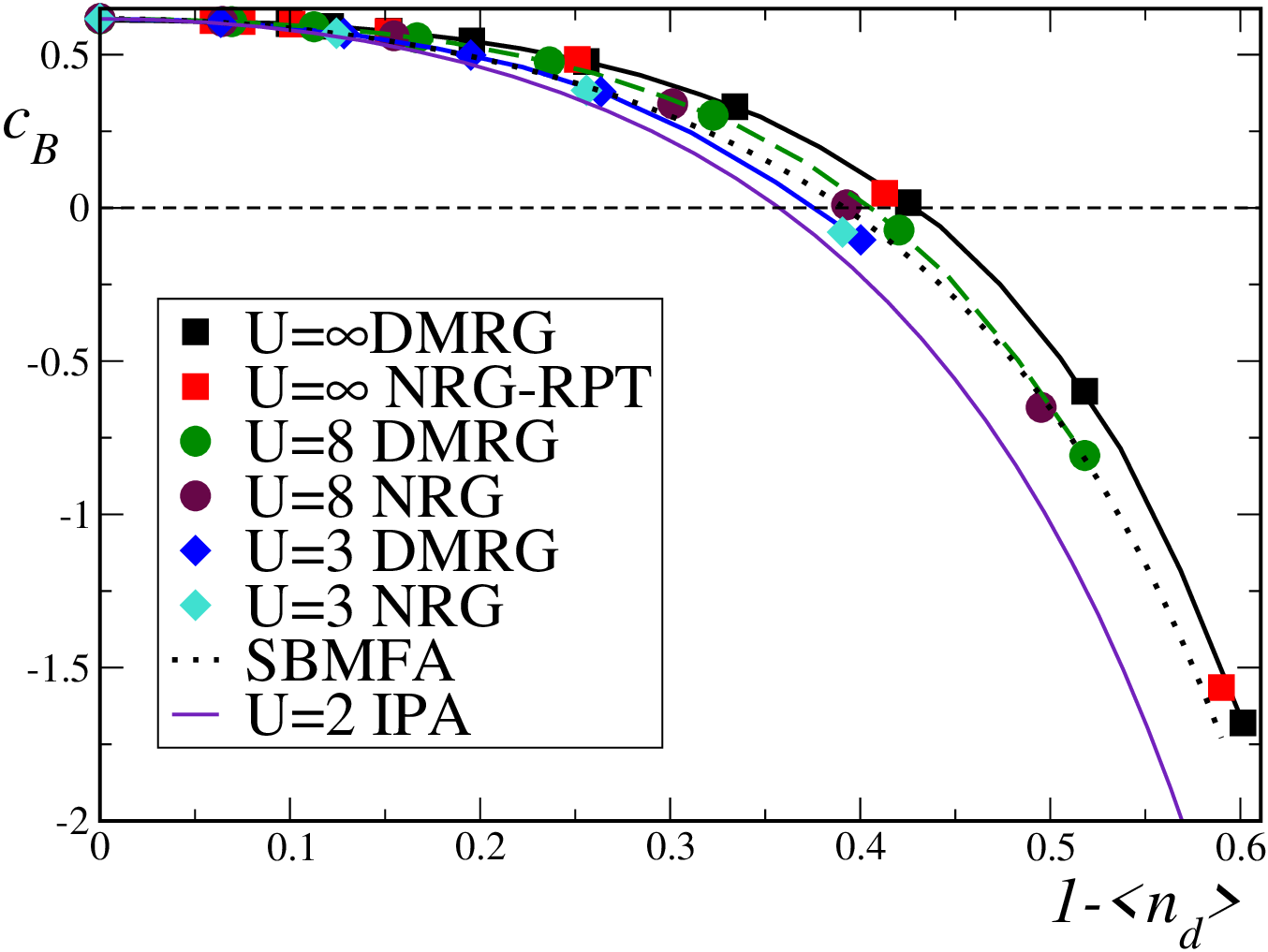}
\caption{(Color online) $c_B$ vs $1 -n_d$ for different values of $U$ obtained by DMRG and NRG+RPT,
and for $U \rightarrow \infty$ using slave bosons in the mean-field approximation.}
\label{cbnd}
\end{figure}
 
In Fig. \ref{cbnd} we represent our main results of $c_B$ as functions of the occupancy
and the corresponding result for $U \rightarrow \infty$ obtained previously using slave
bosons in the mean-field approximation (SBMFA).\cite{cb} 
We also include the results using the interpolative perturbative approximation (see Section \ref{ipa}) for $U=2$. 
In spite of the very different values of
$U$ used, all curves look qualitatively similar. There is a slightly more pronounced decrease 
of $c_B$ vs $1 -n_d$ for smaller values of $U$, $c_B$ changes sign at $n_d \sim 0.64$ for $U=2$ and at $n_d \sim 0.57$ for $U \rightarrow \infty$. 
Curiously, the SBMFA gives values that lie in between those for $U=3 \Delta$ and those for $U=8 \Delta$, 
although it is intended for $U \rightarrow \infty$.

\section{Summary}
\label{sum}

Using different techniques (mainly DMRG and NRG+RPT) we have calculated the coefficient of the 
magnetic-field dependence of the conductance [see Eq. (\ref{cbe})] for  several values of $U$ 
focusing on the strong-coupling limit $U \gg T_K$. As it is known, $c_B=(\pi/4)^2$ in the symmetric
case $E_d=-U/2$, where the occupancy $n_d=1$. As $E_d$ increases or decreases, $c_B$ decreases and changes sign 
in the intermediate valence cases where $E_d^* \sim \epsilon_F$ or $E_d^*+U \sim \epsilon_F$,
where $E_d^*$ is a renormalized energy that 
includes the Haldane shift (which is important for large $U$) and $\epsilon_F$ is the Fermi energy.
In these cases (related by a special electron-hole transformation) the occupancy $n_d$ is near 0.6 
or 1.4 respectively.

We have also provided quantitative results for the ratio of the energy scale of the magnetic-field dependence
of the conductance and the magnetic susceptibility $T_0$ [see Eq. (\ref{cbe})] with the  
half width at half maximum of the spectral density $\Delta_\rho$. For devices in which the coupling to the 
left and right leads are very different, this quantity coincides with the half width at half maximum 
$\Delta_G$ of the zero-bias peak in $G$ times the electric charge $e$  
and is experimentally accessible.\cite{capac,note1} 
For more symmetric devices $\Delta_G/\Delta_\rho$ increases and some values are tabulated in Ref. \onlinecite{capac}.

When $c_B$ is represented as a function of the occupancy $n_d$ for different values of $U$, the curves
$c_B(n_d)$ look qualitatively similar.

\section*{Acknowledgments}

We thank Luis O. Manuel for useful discussions. We are partially supported by CONICET, Argentina. This work was sponsored by
PICT 2013-1045 of the ANPCyT-Argentina, PIP 112-201101-00832, PIP 112-201201-00389, PIP 112-201201-01060 and PIP 112-201201-00273 of CONICET.


\begin{thebibliography}{99}
\bibitem{note} There is a controversy about the non-equilibrium lesser and
greater self-energies,\cite{mu,ng} which we do not address here. These
quantities are discussed in detail in Ref. \onlinecite{ng}, where it is shown
that all expressions satisfy basic Ward identities.
\bibitem{note2} This can be checked using for example perturbation theory
in the Coulomb repulsion $U$ up to second order,\cite{yos,hor} which in the symmetric
case $E_d=-U/2$ is accurate for $U/\Delta \le 2 \pi$.\cite{silver}
\bibitem{note1} While the explicit results of Ref. \onlinecite{capac} were obtained
using the non-crossing approximation, the result that for highly asymmetric devices in the Kondo regime, 
the differential conductance $G=dI/dV_b$ near zero applied bias
voltage $V_b$ reproduces the spectral density of the Kondo peak is based on general
physical arguments and known properties of the Kondo resonance and does not
depend on the approximation.
\end{thebibliography}
\end{document}